\documentclass[aps,prl,reprint,superscriptaddress]{revtex4-2}
\pdfoutput=1

\usepackage[utf8]{inputenc}
\usepackage[T1]{fontenc}
\usepackage{amsmath, amssymb,graphicx,lmodern}
\usepackage{microtype} 
\usepackage[colorlinks=true,linkcolor=blue,citecolor=blue]{hyperref}

\begin{document}

\newcommand{\unit}[1]{\,\mathrm{#1}}

\title{Transport measurements on van der Waals heterostructures under pressure}

\author{B\'alint F\"ul\"op}
\affiliation{Department of Physics, Budapest University of Technology and Economics and Nanoelectronics ``Momentum'' Research Group of the Hungarian Academy of Sciences, Budafoki út 8, 1111 Budapest, Hungary}

\author{Albin M\'arffy}
\affiliation{Department of Physics, Budapest University of Technology and Economics and Nanoelectronics ``Momentum'' Research Group of the Hungarian Academy of Sciences, Budafoki út 8, 1111 Budapest, Hungary}

\author{Endre T\'ov\'ari}
\affiliation{Department of Physics, Budapest University of Technology and Economics and Nanoelectronics ``Momentum'' Research Group of the Hungarian Academy of Sciences, Budafoki út 8, 1111 Budapest, Hungary}

\author{M\'at\'e Kedves}
\affiliation{Department of Physics, Budapest University of Technology and Economics and Nanoelectronics ``Momentum'' Research Group of the Hungarian Academy of Sciences, Budafoki út 8, 1111 Budapest, Hungary}

\author{Simon Zihlmann}
\affiliation{Department of Physics, University of Basel, Klingelbergstrasse 82, CH-4056 Basel, Switzerland}

\author{David Indolese}
\affiliation{Department of Physics, University of Basel, Klingelbergstrasse 82, CH-4056 Basel, Switzerland}

\author{Zolt\'an Kov\'acs-Krausz}
\affiliation{Department of Physics, Budapest University of Technology and Economics and Nanoelectronics ``Momentum'' Research Group of the Hungarian Academy of Sciences, Budafoki út 8, 1111 Budapest, Hungary}

\author{Kenji Watanabe}
\affiliation{Research Center for Functional Materials, National Institute for Materials Science, 1-1 Namiki, Tsukuba 305-0044, Japan}

\author{Takashi Taniguchi}
\affiliation{International Center for Materials Nanoarchitectonics, National Institute for Materials Science,  1-1 Namiki, Tsukuba 305-0044, Japan}

\author{Christian Sch\"onenberger}
\affiliation{Department of Physics, University of Basel, Klingelbergstrasse 82, CH-4056 Basel, Switzerland}

\author{István K\'ezsm\'arki}
\affiliation{Experimental Physics V, Center for Electronic Correlations and Magnetism, University of Augsburg, D-86159 Augsburg, Germany}

\author{P\'eter Makk}
\email{peter.makk@mail.bme.hu}
\affiliation{Department of Physics, Budapest University of Technology and Economics and Nanoelectronics ``Momentum'' Research Group of the Hungarian Academy of Sciences, Budafoki út 8, 1111 Budapest, Hungary}

\author{Szabolcs Csonka}
\email{csonka@mono.eik.bme.hu}
\affiliation{Department of Physics, Budapest University of Technology and Economics and Nanoelectronics ``Momentum'' Research Group of the Hungarian Academy of Sciences, Budafoki út 8, 1111 Budapest, Hungary}

\date{\today}

\begin{abstract}
The interlayer coupling, which has a strong influence on the properties of van der Waals heterostructures, strongly depends on the interlayer distance. 
Although considerable theoretical interest has been demonstrated, experiments exploiting a variable interlayer coupling on nanocircuits are scarce due to the experimental difficulties.
Here, we demonstrate a novel method to tune the interlayer coupling using hydrostatic pressure by incorporating van der Waals heterostructure based nanocircuits in piston-cylinder hydrostatic pressure cells with a dedicated sample holder design.
This technique opens the way to conduct transport measurements on nanodevices under pressure using up to 12 contacts without constraints on the sample at fabrication level.
Using transport measurements, we demonstrate that hexagonal boron nitride capping layer provides a good protection of van der Waals heterostructures from the influence of the pressure medium, and we show experimental evidence of the influence of pressure on the interlayer coupling using weak localization measurements on a TMDC/graphene heterostructure.
\end{abstract}

\maketitle

\section{Introduction}

The discovery of graphene along with other layered crystals led to the emergence of a novel field in material science \cite{Geim2013, Koski2013, Wang2013}.
Van der Waals (vdW) heterostructures, composed of multiple crystal layers with a thickness of a few atomic layers each, exhibit various electronic properties unprecedented in the bulk versions of the component materials. 
The novelty in these heterostructures is twofold.
First, atomically thin layers show novel electronic properties compared to their bulk counterparts, such as the massless Dirac fermion nature of electrons in graphene \cite{Novoselov2005}, a direct-to-indirect semiconductor band gap transition in MoS$_2$ with decreasing number of layers\cite{Mak2010}, the magnetic structure of CrI$_3$ that depends on the number of layers \cite{Huang2017}, or the appearance of a topological insulating phase in single layer WTe$_2$.\cite{Qian2014, Fei2017, Tang2017, Wu2018, Shi2019, Zhao2020}.
Second, combining these layers into heterostructures leads to overlapping atomic orbitals in the neighbouring layers, which may cause drastic changes in the electronic structure of the system.
This often originates from the different materials of the components, e.g.\ the emergence of inherited proximity spin orbit coupling (SOC) \cite{Gmitra2016, Wang2015, Wang2016, Benitez2018, Zihlmann2018, Ghiasi2017, Wakamura2019, Fulop2021wal}, ferromagnetic ordering in graphene\cite{Leutenantsmeyer2017, Karpiak2019, Wang2015a, Ghiasi2020} or in transition metal dichalchogenides \cite{Zhong2017,Zhong2020}.
A moiré superlattice can also emerge if the component layers have unit cells of almost the same size, as in the case of graphene with hexagonal boron nitride (hBN), leading to the appearance of secondary Dirac points, Hofstadter butterflies and Brown-Zak oscillations in transport \cite{Dean2013, Ponomarenko2013, Hunt2013, KrishnaKumar2017, KrishnaKumar2018, Ribeiro-Palau2018, Wang2019}.
Moreover, for certain rotation angles, so called magic angles, flat bands and novel phases of material can form as e.g.\ in the case of twisted bilayer graphene \cite{cao2018a, cao2018b}.

\begin{figure}
\centering
\includegraphics[width=8.5cm]{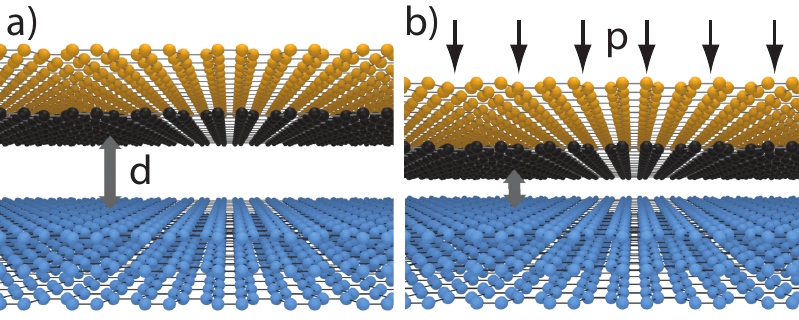}
\caption{
Schematic illustration of the effect of hydrostatic pressure on an arbitrary van der Waals heterostructure. 
By applying an external pressure $p$, the heterostructure compresses, thus changing the distance $d$ and the interaction strength between the component layers.
}
\label{fig:vdw}
\end{figure}

The interlayer coupling strongly depends on the interlayer distance, which is determined by the vdW interaction that keeps the layers together.
However, by applying hydrostatic pressure on these heterostructures (see fig.\,\ref{fig:vdw}), the interlayer coupling can be tuned, which strongly influences the electronic structure and thereby the physical properties of these heterostructures.
The potential interest of tuning this parameter is demonstrated in various theoretical works \cite{Weng2014, Gmitra2016, Munoz2016, Fan2017, Nam2017, Zhang2017, Guinea2019, Chebrolu2019, Lin2020}, however, several technological challenges are to be solved for the realization of transport measurements on nanodevices under pressure.

Hydrostatic pressure cells have been widely used in solid state physics on bulk samples.
Changing the pressure can not only tune the wavefuction overlap but can lead to structural changes in the crystal, as e.g.\ in 2H$\mathrm{_c}$-MoS$_2$, where  layer sliding structural transition and metal-insulator transition take place \cite{Dave2004, Aksoy2006, Aksoy2008, Hromadova2013, riflikova2014}. 
Pressure is an experimental knob to modify different correlated states and induce quantum phase transitions from charge density wave to superconductivity \cite{Zamborszky2001, kezsmarki2007, kusmartseva2009, vangennep2016}, or from paramagnetism to ferromagnetism \cite{ forro2000, kezsmarki2001, kezsmarki2005, csontos2005}.   

All the experimental works on transport measurements with pressure manipulation cited above used samples that remained in the macroscopic size range (0.1\,mm), even in case of carbon nanotubes, the sample was on a buckypaper of macroscopic size \cite{barisic2002}. 

Meanwhile, the dimensions of vdW heterostructure based nanocircuits are microscopic and conventionally a very different measurement method is used to carry out transport measurements, often involving wire bonding on standard ceramic chip carriers or circuit boards.
The spatial dimensions of the conventional solutions are considerably larger than the available space in a common hydrostatic pressure cell, which is limited to technological constraints in order to achieve high hydrostatic pressure values without damaging the cell components.
Therefore, the standard measurement techniques of nanocircuits do not work in a hydrostatic pressure environment, and special setups are needed.

The first pioneering papers about such experiments already showed the potential of this technique, where they managed to tune the superconducting critical temperature of a twisted bilayer graphene structure or change the magnetization alignment of a thin magnetic structure \cite{Yankowitz2018, Yankowitz2019, Song2019}.
In these studies, the vdW nanocircuits were fixed and electrically connected using silver paint directly to the pressure cell feed-through wires without any intermediate structure.
Although it can work, this technique has several limitations:
it requires very accurate manual work, and electrostatic discharge protection is not fulfilled during silver paint contacting. 
Furthermore, precise orientation of the chip inside the cell also remains challenging, which is required as the proper relative orientation of the vdW heterostructures to the magnet axes play an important role in many measurements \cite{Leutenantsmeyer2018, Xu2018}.
Therefore, a systematic, safe and easy method to incorporate nanocircuit measurements into a hydrostatic pressure cell environment is still missing.
In this paper, we present such a new method and a dedicated sample holder for measurements of nanocircuits using wire-bonding up to 12 electric lines with controlled and reproducible positioning of the chip on the cell plug.
Our method is fully compatible with the standard sample fabrication methods and does not require specific design at fabrication level unlike previous studies \cite{Yankowitz2018, Yankowitz2019}, which makes hydrostatic measurements possible on any sample below ca.\ 3\,mm lateral size, equipped with bonding pads.

In the next section, after detailing the specific technical requirements of transport measurements on a nanocircuit under pressure, we introduce our dedicated sample holder head and its working mechanism. 
Then, we present two different examples how pressure can be applied on van der Waals heterostructure based nanocircuits. 
The first measurement demonstrates that an hBN cover layer provides an ideal isolation between the sensitive 2D electron system and the pressure transfer medium. 
The second one gives evidence that the interlayer coupling can be changed significantly in graphene/TMDC heterostructures by the pressure generated in our setup.

\begin{figure*}
\includegraphics{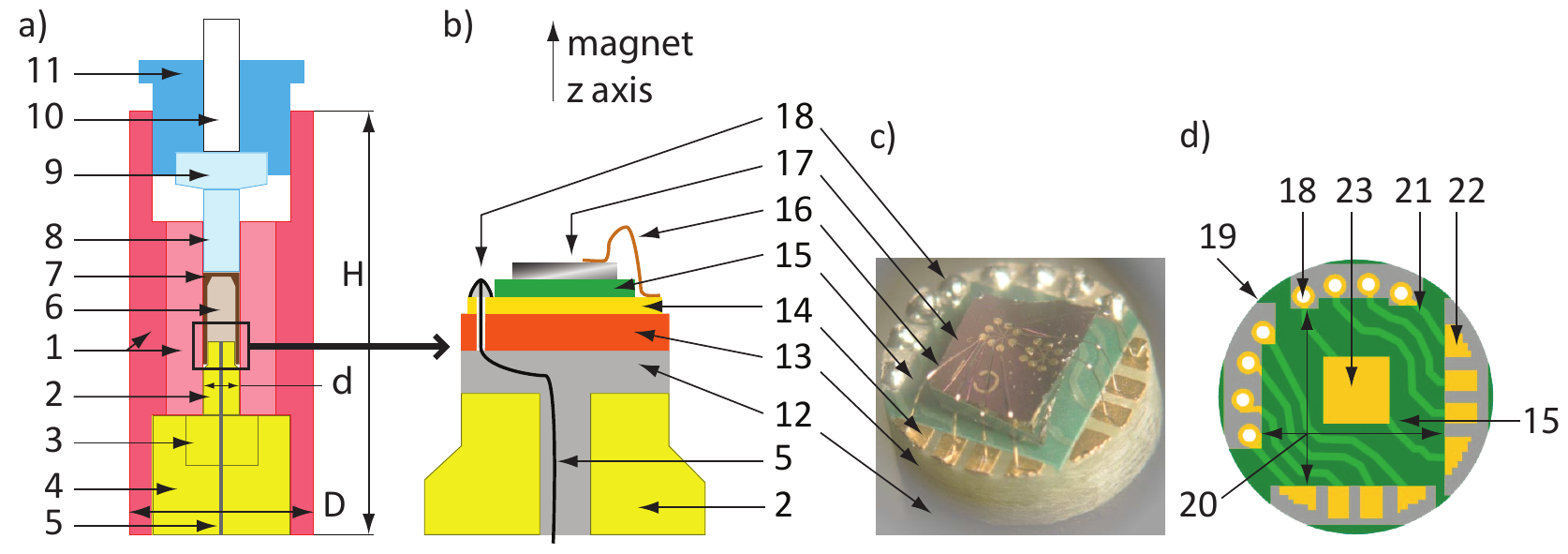}
\caption{
(a)
Schematic side view of a clamp-type two-layer pressure cell without the detailed sample holder head. 
The cell dimensions are $H = 62\unit{mm}$, $D=25\unit{mm}$, and $d=6.5\unit{mm}$.
1. NiCrAl/CuBe double layered cell wall.
2.\,CuBe plug.
3.\,WC plug backup.
4.\,CuBe lower screw.
5.\,Electric feed-through, marked by grey across the yellow parts.
6.\,High pressure volume (HPV), filled with pressure medium.
7.\,Teflon cup.
8.\,WC piston.
9.\,WC piston backup.
10.\,WC push rod.
11.\,CuBe upper screw.
Additional sealing rings (not drawn) prevent oil leak from the HPV\@.
(b)
Detailed schematic side view of the sample holder head, as marked by the black rectangle on panel\,a.
(c)
Corresponding bird's-eye view of the top of the sample holder head with a chip attached on it and bonded to the electric contacts.
Marks 2.\ and 5.\ are the same as in panel\,a.
5.\,Feed-through diameter varies between 0.7\,mm and 1.0\,mm between plug instances.
12.\,Epoxy bedding.
13.\,PCB core layer.
14.\,PCB top Cu layer with Ni/Au finish.
15.\,PCB solder stop layer.
16.\,Bonding wires connecting the PCB bonding pad and the lithographed contact pads of the chip.
17.\,Nanocircuit chip with the heterostructure on its surface.
18.\,Non-plated through-holes (NPTHs) and soldered electric feed-through lines.
(d)
Schematic top view of the PCB design (8-line version). 
19.\,The useful PCB diameter is $5.1\unit{mm}$. 
20.\,Central area (ca.\ 3.5\,mm$\times$3.5\,mm) where the chip is placed. Maximal chip size ca.\ 3\,mm$\times$3\,mm.
21.\,PCB traces connecting the soldered feed-through lines to the bonding pads.
22.\,Bonding pads.
23.\,Au backplate to contact the chip backside, if needed.
}
\label{fig:cell}
\end{figure*}

\section{Sample holder design}

Piston-cylinder pressure cells (see fig.\,\ref{fig:cell}a and the \textit{methods} for a detailed description) have been widely used before to measure transport through macroscopic samples. 
Contacting electric wires to macroscopic samples using silver paste for two or four-terminal measurements is feasible by hand, although it requires operator experience.
Meanwhile, transport measurements on nanocircuits built on conventional Si/SiO$_2$ substrates can be far more challenging.

First, measurements of a complicated sample may involve several contacts on the chip surface, occasionally contacting the doped Si substrate on the chip bottom as a global back gate as well. 
There might also be multiple equally interesting devices on the same chip, often resulting in more than 20 contacts on a single chip.
Second, contact pads usually created by electron lithography on these nanocircuits vary in size typically between 20~\textmu{}m and 200~\textmu{}m with a spacing between them of the same length scale.
Contacting such electronic contact pads by hand using silver paste is hardly feasible.
It is possible to design the chip with enlarged contact pads and placed far from each other, (thus limiting the number of available contacts) \cite{Yankowitz2018,Yankowitz2019}, but even in this case, application of silver paste puts the sample at considerable risk of accidentally contaminating the chip surface, or of uncontrolled electrostatic discharge (ESD).
This dangerous situation is repeated if the sample needs to be taken off the sample holder plug temporarily and placed back later (e.g.\ to conduct an auxiliary measurement on it), needing re-connection of the lithographed contacts.
Third, vdW heterostructures are anisotropic samples by design. 
Therefore, chip orientation relative to the cryostat magnet axes plays an important role and needs to be controlled and maintained during the experiments, which can last several days with multiple thermal cycles as well.
Orientation reproducibility is also required in case of a repeated installation of the sample on the plug.

We developed a dedicated sample holder head to satisfy these requirements.
The head consists of the plug, the electric feed-through wires embedded in an epoxy filler, and a printed circuit board (PCB).
The sample is fixed on the PCB using conducting or insulating double sided tape and is contacted using wire-bonding to the PCB contact pads.
Using the standard wire-bonding technology, one can use nanocircuits without any special modification at lithography level and with several bonding pads of the usual size.

A schematic view of the sample holder is depicted in fig.\,\ref{fig:cell}b, whereas in fig.\,\ref{fig:cell}c a bird's-eye view photograph is shown.
36\,AWG (american wire gauge, diameter $D \approx130$\,\textmu{}m) phosphor-bronze electrical wires (5) are used to connect the PCB to external connectors.
Stycast 2850FT epoxy filler with catalyst 9 (12) is used to seal the plug through-hole around the feed-through wires.
Its role is not only to seal, but also to support the PCB (layers 4-6) mechanically, finally, to fill the space below it to avoid air bubble formation that can lead to mechanical shock waves when the bubbles collapse during pressurization.
The epoxy is filled in the plug through-hole but not in the hole of the plug backup or of the lower screw.
The electric feed-through wires penetrate the PCB core layer (13) via non-plated through-holes (NPTHs, 18) and are soldered to traces (14) on the top of the PCB that lead to bonding pads under a solder stop layer (15) used for isolation from the possibly conducting backside of the chip (17).
The chip contacts are connected via wire-bonding (16).
The PCB surface is levelled perpendicular to the plug axis, thus securing the chip in a well defined orientation with respect to the cryostat magnet axes.
Levelling the PCB surface is a delicate handwork at the sample holder fabrication time, but using the already soldered wires' rigidity, it can be done with less than 5$^\circ$ tilt.
The epoxy filler is applied after this step.

We developed three different versions of the PCB design with 8, 10, and 12 contacts, of which the 8-line version is shown in fig.\,\ref{fig:cell}d. 
The useful diameter of the PCB is limited by the inner diameter of the cell and the thickness of the Teflon cup to 5.1\,mm (19).
A clearance of $0.25\unit{mm}$ is also added along the perimeter for PCB manufacturing reasons which is scraped away as the last step of the sample holder production.
The central part (20) of the PCB supports the chip itself on an area of $3.5 \times 3.5 \unit{mm^2}$ (8-line variant).
The finite space requirement of the wire bonding procedure and a tolerance of positioning the chip manually by tweezers altogether limits the chip size to ca. $3 \times 3 \unit{mm^2}$ without the risk of shorting any of the contacts or obscuring the bonding pads.
The feed-through wires penetrate the PCB core layer through the NPTHs lined up at the top and left side of the PCB (18).
Traces (21) link the NPTHs to the bonding pads (22) under the solder stop layer (15) to avoid shorts to the chip substrate.
Special care should be taken at the sample holder fabrication time to avoid solder on the bonding pads, which makes the bonding wires unable to stick on the pad surface.
Therefore, the bonding pads, covered by Ni/Au finish and enlarged as much as possible to maximize reusability, are placed on the opposite sides to the NPTHs.
One trace leads through a back plate region in the central part of the PCB (20), which is not covered by the solder stop.
This part can serve as a connection to the chip substrate used as a back gate if the chip is fixed with a carbon tape, or can be completely ignored if an insulating tape is used.
In the latter case, this line can also serve as a regular connection via its bonding pad.
Using adhesive tapes instead of the commonly used silver paste or silver epoxy glue has several advantages: the tapes offer the possibility to choose between conducting and isolating fixing layer, it is much faster than applying a paste due to the lack of drying, and it is much cleaner, i.e.\ does not contaminate either the PCB surface or the chip.

Another advantage of the dedicated PCB design is that as long as not exposed to the pressure medium, this sample holder head works just as any regular measurement setup of nanocircuits and can be a substitute of it.
In addition, it is possible to mass fabricate the sample holder head prior to experiments and if multiple sets of the plug, plug backup and lower screw are available, parallel investigation of different samples is also feasible, reducing the need of de-connecting or re-connecting samples.
The other variants of our PCB design, offering 10 and 12 electric connections, limit the sample size to ca.\ 2.5\,mm$\times$3.0\,mm and 1.7\,mm$\times$1.7\,mm, respectively.

\section{Measurement results}

\begin{figure}
\includegraphics{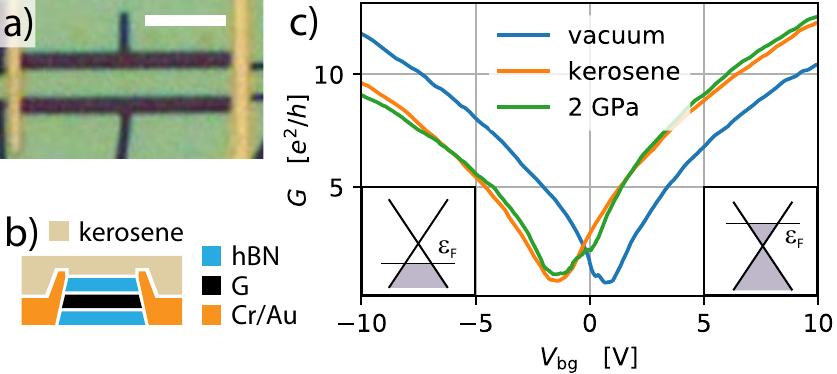}
\caption{
Measurements on device A.
(a)
Optical microgaph.
The green stripe is the heterostructure, the electrical contacts are on the two sides and the dark blue stripes are the surface of the bare SiO$_2$ chip, where etching shaped the heterostructure.
The current in the detailed two-terminal measurement flows along the horizontal long middle section of dimensions $W=1.7$\,\textmu{}m and $L=12.7$\,\textmu{}m.
The scalebar (white) is 5\,\textmu{}m.
(b)
Schematic side view of the stack as it is placed on the SiO$_2$ surface.
A single layer graphene (G) sheet is encapsulated between hexagonal boron nitride (hBN) layers and contacted to Cr/Au electrodes via standard line-contacts.
(c)
Comparison of two-terminal conductance vs.\ back gate voltage measurements at 4\,K\@.
Vacuum: the sample has never been exposed to kerosene. 
kerosene: in kerosene environment, without additional pressure. 
2.5\,GPa: in kerosene environment, under 2.5\,GPa pressure. 
The top hBN flake protects the graphene layer and its electronic contacts from contamination in the kerosene environment.
Left (right) inset:
schematic view of the Fermi energy in the band structure of graphene as it is tuned in the valence (conduction) band by the gate voltage.
}
\label{fig:gatesweeps}
\end{figure}

Due to its monolayer nature, the electronic quality of a gaphene flake is largely affected by the presence of charged impurities on its surface \cite{CastroNeto2009}.
This contamination induces a rough electrostatic potential that leads to scattering mechanisms that reduce the charge carrier mobility and the mean free path in the sample, which decreases the visibility of many transport phenomena of interest.
Such contamination is expected to occur when the graphene flake is exposed to the kerosene used as pressure medium in the experiment, since it is an organic liquid with many components of various chemical structure.
Indeed, our previous experience shows that exposing a single graphene layer to kerosene degrades its electronic quality even to a point where the charge neutrality point (CNP) is not visible anymore in the transport measurements.
We studied and hereby demonstrate that encapsulation with hexagonal boron nitride (hBN) flakes of thickness in the order of 10\,nm can protect the graphene layer from the pressure medium, and in such a heterostructure, the presence of kerosene does not affect substantially either the contact resistances or the charge carrier mobility measured in two-terminal measurements.

The studied devices were fabricated on a Si/SiO$_2$ substrate using the common dry stacking assembly method \cite{Zomer2014}. 
The shape of the samples were defined using reactive ion etching and Cr/Au side contacts were fabricated in a separate lithography step \cite{Wang2013}.
Two terminal measurements were performed at $f=177\unit{Hz}$ with $V_\mathrm{AC} = 100$\,\textmu{}V excitation voltage using standard lock-in technique and external low-noise current amplifier. 
The doped Si substrate is used as a global backgate electrode.

Device A consists of a single layer of graphene encapsulated in multilayer top and bottom hBN flakes, see fig.\,\ref{fig:gatesweeps}a-b and was shaped into a stripe of dimensions $W = 1.7$\,\textmu{}m and $L=12.7$\,\textmu{}m.
In fig.\,\ref{fig:gatesweeps}c, we present two-terminal conductance measurements at 4\,K temperature in different environments as a function of the gate voltage: first, before any exposure to kerosene, then in kerosene without applying external pressure, and finally applying 2.5\,GPa pressure.
The charge neutrality point (CNP) is shifted slightly, from $0.7\unit{V}$ to $-1.5\unit{V}$ after the exposure to kerosene, but did not change during pressurization.
For the extraction of the charge carrier mobility and the series resistance of the contacts, a curve fitting was performed using a parallel plate capacitor model to determine the charge carrier density.
The mobility stayed within a 10\% range between the cases, reaching $55,000\unit{cm^2/Vs}$ and $50,000\unit{cm^2/Vs}$ for electrons and holes, respectively.
The contact resistances remained in the $220-340\unit{\Omega\cdot}$\textmu{}m range without any systematic trend.
As the variation of these parameters between subsequent thermal cycles of the same heterostructure is usually in the same order of magnitude, we conclude that the presence of the kerosene does not change the mobility of the sample, nor the quality of the edge contacts. 
We have performed similar measurements on various devices with different van der Waals heterostructures using hBN protection layer on the top and we found that the protection property of hBN flakes is generic. 

\begin{figure}
\includegraphics{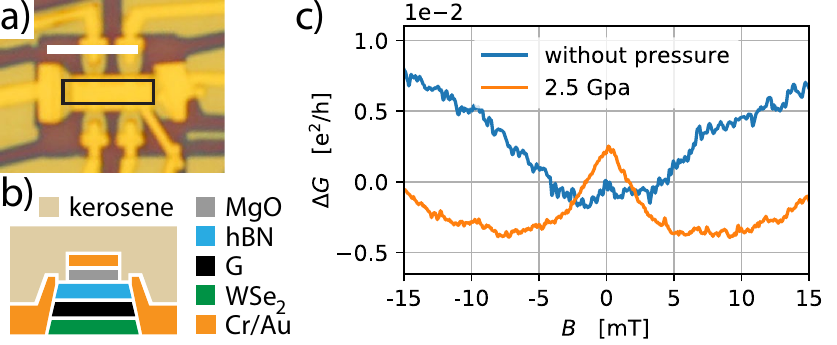}
\caption{
Measurements on device B.
(a)
Optical image of the device.
The current in the detailed measurement flows along the longest segment, highlighted by the black line.
The scalebar is 5\,\textmu{}m.
(b)
Schematic side view of the stack as it is placed on the SiO$_2$ surface.
The graphene layer (G) is placed on a multilayer WSe$_2$ flake and covered by a multilayer hBN.
An additional topgate covers partially the heterostructure separated by an insulating MgO layer, but it was not used in the presented measurements.
(c)
Two-terminal magneto-conductance curves in kerosene environment under ambient pressure and 2.5\,GPa, the latter shifted by $+2\cdot10^{-3} e^2/h$ for clarity.
A clear enhancement of weak antilocalization is visible indicating an enhancement of the proximity-induced spin orbit coupling in the graphene layer.
}
\label{fig:WAL}
\end{figure}

As can be seen from fig.\,\ref{fig:gatesweeps}, the transport properties of single layer graphene are not substantially changed under hydrostatic pressure. 
This changes markedly for multi-layer systems. 
One of the large potentials of the experiments under hydrostatic pressure lies in tuning interfacial proximity interactions in vdW heterostructures. 
After demonstrating the possibility of protecting the heterostructures in the hydrostatic pressure cell environment, we show that the applied pressure can result in an observable change in the transport properties of a graphene sample coupled to a WSe$_2$ crystal.
On fig.\,\ref{fig:WAL}a the optical micrograph of sample B is shown, where the measured segment of $W = 1$\,\textmu{}m and $L=5$\,\textmu{}m is highlighted by a black line. 
The schematic side view of the sample is shown in fig.\,\ref{fig:WAL}b, which is composed of a bottom multilayer WSe$_2$ flake, a single graphene flake and a multilayer hBN protecting flake on the top.
The device is also equipped with a topgate, separated from the heterostructure and its contacts by an insulating MgO layer, but this was not used in the experiments presented here.
In such TMDC/graphene heterostructures, it was found that the TMDCs can induce a spin orbit interaction in graphene via proximity effect \cite{Avsar2014, Wang2015, Wang2016, Yang2016, Ghiasi2017, Yang2017, Benitez2018, Zihlmann2018, Ringer2018, Island2019, Wakamura2019, Amann2012.05718}. 
The strength of the spin orbit interaction depends on the layer distance, which is expected to be tuned by hydrostatic pressure.
One fingerprint of the spin orbit interaction in transport measurements is the presence of weak anti-localization (WAL) feature in the magneto-conductance (MC) curves, i.e.\ the change in the conductance vs.\ the out of plane magnetic field \cite{IhnBook2004}.

Fig.\,\ref{fig:WAL}c shows MC curves in kerosene environment at 4\,K temperature, one with no pressure, the other with 2.5\,GPa pressure.
Both show negative MC which corresponds to the WAL effect for very small magnetic fields, but the signal amplitude in the latter case is much larger and the peak is much wider, which are indications of the enhancement of the induced spin orbit coupling.

Whereas the formation of the WAL peak might be related to changes in the intervalley scattering time due to the Berry phase in graphene, we suspect that it is rather related to the increased spin orbit coupling strength in graphene. 
The detailed analysis goes well beyond the scope of this manuscript and is discussed elsewhere \cite{Fulop2021wal_arxiv}.

The sample was not designed specifically to the needs of the pressure cell, however, it could be incorporated into our setup. 
Three devices were formed on a single chip, each with a Hall bar shape with several electrical contacts and local top gates, with 23 contacts in total on the chip surface.
A contact pad on the chip surface was $200\times150$\,\textmu{}m$^2$ with a spacing of 30\,\textmu{}m.
This is a size scale that is really challenging to bond using silver paint without accidentally connecting neighbouring contact pads or even contaminating the nanocircuit itself.
The bounding box of the patterned area extended to $2\times2\unit{mm}$ which needs precise dicing of the chip in order to fit on our sample holder.
We note that the performed measurements included several thermal cycles with multiple re-bonding of the wires and lasted for several weeks in total without any problem in the sample holder head.

\section{Conclusion}
In this work we presented a novel method to tune the interlayer coupling in vdW heterostructures post fabrication using hydrostatic pressure.
This method enables electronic transport measurements on vdW heterostructures while adding the possibility of tuning the interlayer coupling in them at measurement time.
This method allows controlled and reversible fixation of the chip and up to 12 wire-bonding connections to it, which makes hydrostatic pressure studies a lot easier than existing alternatives. 
We showed experimental evidence of the applicability of hBN crystal flakes as a top layer in order to protect the underlying layers from the pressure medium.
Changing interlayer coupling has a measurable effect on the electrical transport of a sensitive heterostructure, as has been demonstrated by weak localization measurements on a graphene/WSe$_2$ heterostructure.
Thus, our study provides a new experimental knob which can be used to tune vdW heterostructure based nanodevices and opens the way towards the highly anticipated pressure studies not only for vdW heterostructures, but also for any other nanocircuit-based physical systems.

\section{Authors contribution}

B.F, A. M. , Z. K-K. and E.T developped the pressure cell setup with the guidance of Sz.Cs and I. K.
The samples were fabricated by S.Z., D.I., M.K. with the support of P.M. 
B. F. and A. M. did the measurements.
The manuscript was written by B.F., P.M and Sz.Cs. 
All authors discussed on the manuscript.
K.W. and T.T. grew the hBN crystals. 
The project was guided by Sz.Cs., P.M and C.S.

\section{Acknowledgements}

This work acknowledges support from the Topograph FlagERA network, the OTKA FK- 123894 grants, the Swiss Nanoscience Institute (SNI), the ERC project Top-Supra (787414), the Swiss National Science Foundation, the Swiss NCCR QSIT.  
P.M. and E.T. received funding from Bolyai Fellowship. 
This research was supportedby the Ministry of Innovation and Technology and the National Research, Development and Innovation Office within the Quantum Information National Laboratory of Hungary and by the Quantum Technology National Excellence Program (Project Nr. 2017-1.2.1-NKP-2017-00001). 
K.W. and T.T. acknowledge support from the Elemental Strategy Initiative conducted by the MEXT, Japan, Grant Number JPMXP0112101001, JSPS KAKENHI Grant Numbers JP20H00354 and the CREST(JPMJCR15F3), JST.

\section{methods}

Although the method presented in this paper works well with any type of piston-cylinder pressure cell, we present in the following, as an example, the general build-up and working principle of one specific cell that we used, made by C\&T Factory, Japan (see fig.\,\ref{fig:cell}a for a schematic side view). 

\textbf{Working principle.}
The sample is placed inside a cylindrical high pressure volume (HPV) at the core of the cell, which is filled with a pressure medium, then compressed in a hydraulic press.
After clamping the cell at the target pressure, it can be removed from the press and used in any transport measurement setup, e.g.\ attached to a cryostat dipstick and cooled down to cryogenic temperature.
At the end of the measurements, to release the pressure or to set a different one, the cell should be warmed up and placed in the hydraulic press again for a reverse process.

The side of the HPV cylinder is supported by the shell of the pressure cell, whereas the top and bottom ends are closed by a piston and a plug element, respectively, both clamped by two screws.
Transport experiments can be done on the sample via an electric feed-through of the plug.
Our cell has an outer diameter of $D = 25\unit{mm}$ and a height $H = 62\unit{mm}$, which is small enough to fit in the bore of many magnets used in cryogenic systems allowing measurements under magnetic field, and also in our VTI measurement setup, thus offering a temperature range of 1.5\,K--300\,K.

\textbf{Build-up of the cell.}
The cell wall is composed of two cylindrical shells made of NiCrAl and CuBe, respectively, see mark (1) in fig.\,\ref{fig:cell}a.
The HPV (6) is situated in the central part of the device in a Teflon cup (7), which opens at the lower end and encloses the pressure medium along with the sample.
The volume is closed on the lower end of the Teflon cup by the sample holder plug (2), the plug backup (3) and clamped by the lower screw (4).
Our special sample holder, discussed below, is fabricated on the top of the plug (2), and the sample is placed on the top of the sample holder, as depicted in fig.\,\ref{fig:cell}b.
The yellow parts are secured in place during setup assembly prior to pressurization. 
Transport measurement inside the cell is carried out via the electric feed-through (5), marked by grey across the yellow parts.
At the top end of the HPV, the piston (8) and the piston backup (9) move downwards in the cell during pressurization, as the push rod (10) transfers the load from an external pump on the Teflon cup and the HPV compresses.
To fix the pressure inside the HPV at the end of the pressurization process, the upper screw (11) is tightened.
Finally, this screw takes the load as the pressure in the external press is released.
For sealing, a CuBe obturator ring is placed between the Teflon cup (7) and the piston (8), whereas on the lower side, a CuZn obturator is supported by a narrow shoulder on the plug at the opening of the Teflon cup. 
The choice of a softer material for the lower obturator ring aims to avoid it sticking inside the cell during disassembly after releasing the pressure, because sticking can lead to the break of the sample holder head and prevents its potential reusability.

\textbf{Pressure medium.}
The HPV inside the Teflon cup (6 in fig.\,\ref{fig:cell}a) is filled by a pressure medium, usually an organic compound. 
We use the Daphne 7373 kerosene compound produced by Idemitsu Co., Japan. 
This type of oil has several properties that make it a good choice as a pressure medium, as studied before \cite{Murata1997,Yokogawa2007}. 
First, it solidifies at room temperature at a relatively high, 2.2\,GPa pressure.
Up to this pressure, hydrostaticity (isotropy and uniformity) of the pressure is assured, but the sample has to be at room temperature.
Second, the pressure reduction from room temperature to cryogenic temperatures is low, around 0.15\,GPa, independently of the room temperature pressure above 0.5\,GPa.
Third, the pressure reduction occurs continuously and gently along the cooling cycle, including the freezing of the oil. 
This helps to avoid problems due to the mechanical shock occurring at a sudden freezing, which could potentially break the bonding wires, the PCB, or the sample itself as well.
To further reduce the chance of mechanical shocks, the temperature ramp rate is limited to 3\,K/min above 150\,K, where the freezing occurs.

\bibliography{D:/in_this_folder}

\begin{thebibliography}{10}

\bibitem{Geim2013}
A.~K. Geim and I.~V. Grigorieva, Van der Waals heterostructures {\em Nature},
  vol.~499, pp.~419--425, July 2013.

\bibitem{Koski2013}
K.~J. Koski and Y.~Cui, The New Skinny in Two-Dimensional Nanomaterials {\em
  ACS Nano}, vol.~7, pp.~3739--3743, May 2013.

\bibitem{Wang2013}
L.~Wang, I.~Meric, P.~Y. Huang, Q.~Gao, Y.~Gao, H.~Tran, T.~Taniguchi,
  K.~Watanabe, L.~M. Campos, D.~A. Muller, J.~Guo, P.~Kim, J.~Hone, K.~L.
  Shepard, and C.~R. Dean, One-Dimensional Electrical Contact to a
  Two-Dimensional Material {\em Science}, vol.~342, no.~6158, pp.~614--617,
  2013.

\bibitem{Novoselov2005}
K.~S. Novoselov, A.~K. Geim, S.~V. Morozov, D.~Jiang, M.~I. Katsnelson, I.~V.
  Grigorieva, S.~V. Dubonos, and A.~A. Firsov, Two-dimensional gas of massless
  Dirac fermions in graphene {\em Nature}, vol.~438, pp.~197--200, Nov. 2005.

\bibitem{Mak2010}
K.~F. Mak, C.~Lee, J.~Hone, J.~Shan, and T.~F. Heinz, Atomically Thin
  ${\mathrm{MoS}}_{2}$: A New Direct-Gap Semiconductor {\em Phys. Rev. Lett.},
  vol.~105, p.~136805, Sept. 2010.

\bibitem{Huang2017}
B.~Huang, G.~Clark, E.~Navarro-Moratalla, D.~R. Klein, R.~Cheng, K.~L. Seyler,
  D.~Zhong, E.~Schmidgall, M.~A. McGuire, D.~H. Cobden, W.~Yao, D.~Xiao,
  P.~Jarillo-Herrero, and X.~Xu, Layer-dependent ferromagnetism in a van der
  Waals crystal down to the monolayer limit {\em Nature}, vol.~546,
  pp.~270--273, June 2017.

\bibitem{Qian2014}
X.~Qian, J.~Liu, L.~Fu, and J.~Li, Quantum spin Hall effect in two-dimensional
  transition metal dichalcogenides {\em Science}, vol.~346, p.~1344, Dec. 2014.

\bibitem{Fei2017}
Z.~Fei, T.~Palomaki, S.~Wu, W.~Zhao, X.~Cai, B.~Sun, P.~Nguyen, J.~Finney,
  X.~Xu, and D.~H. Cobden, Edge conduction in monolayer WTe2 {\em Nature
  Physics}, vol.~13, pp.~677--682, July 2017.

\bibitem{Tang2017}
S.~Tang, C.~Zhang, D.~Wong, Z.~Pedramrazi, H.-Z. Tsai, C.~Jia, B.~Moritz,
  M.~Claassen, H.~Ryu, S.~Kahn, J.~Jiang, H.~Yan, M.~Hashimoto, D.~Lu, R.~G.
  Moore, C.-C. Hwang, C.~Hwang, Z.~Hussain, Y.~Chen, M.~M. Ugeda, Z.~Liu,
  X.~Xie, T.~P. Devereaux, M.~F. Crommie, S.-K. Mo, and Z.-X. Shen, Quantum
  spin Hall state in monolayer 1T'-WTe2 {\em Nature Physics}, vol.~13,
  pp.~683--687, July 2017.

\bibitem{Wu2018}
S.~Wu, V.~Fatemi, Q.~D. Gibson, K.~Watanabe, T.~Taniguchi, R.~J. Cava, and
  P.~Jarillo-Herrero, Observation of the quantum spin Hall effect up to 100
  kelvin in a monolayer crystal {\em Science}, vol.~359, p.~76, Jan. 2018.

\bibitem{Shi2019}
Y.~Shi, J.~Kahn, B.~Niu, Z.~Fei, B.~Sun, X.~Cai, B.~A. Francisco, D.~Wu, Z.-X.
  Shen, X.~Xu, D.~H. Cobden, and Y.-T. Cui, Imaging quantum spin Hall edges in
  monolayer WTe\&lt;sub\&gt;2\&lt;/sub\&gt; {\em Sci Adv}, vol.~5, p.~eaat8799,
  Feb. 2019.

\bibitem{Zhao2020}
W.~Zhao, Z.~Fei, T.~Song, H.~K. Choi, T.~Palomaki, B.~Sun, P.~Malinowski, M.~A.
  McGuire, J.-H. Chu, X.~Xu, and D.~H. Cobden, Magnetic proximity and
  nonreciprocal current switching in a monolayer WTe2 helical edge {\em Nature
  Materials}, vol.~19, pp.~503--507, May 2020.

\bibitem{Gmitra2016}
M.~Gmitra, D.~Kochan, P.~H\"ogl, and J.~Fabian, Trivial and inverted Dirac
  bands and the emergence of quantum spin Hall states in graphene on
  transition-metal dichalcogenides {\em Phys. Rev. B}, vol.~93, p.~155104, Apr
  2016.

\bibitem{Wang2015}
Z.~Wang, D.-K. Ki, H.~Chen, H.~Berger, A.~H. MacDonald, and A.~F. Morpurgo,
  Strong interface-induced spin-orbit interaction in graphene on WS2 {\em
  Nature Communications}, vol.~6, p.~8339, Sept. 2015.

\bibitem{Wang2016}
Z.~Wang, D.-K. Ki, J.~Y. Khoo, D.~Mauro, H.~Berger, L.~S. Levitov, and A.~F.
  Morpurgo, Origin and Magnitude of `Designer' Spin-Orbit Interaction in
  Graphene on Semiconducting Transition Metal Dichalcogenides {\em Phys. Rev.
  X}, vol.~6, p.~041020, Oct. 2016.

\bibitem{Benitez2018}
L.~A. Benítez, J.~F. Sierra, W.~Savero~Torres, A.~Arrighi, F.~Bonell, M.~V.
  Costache, and S.~O. Valenzuela, Strongly anisotropic spin relaxation in
  graphene-transition metal dichalcogenide heterostructures at room temperature
  {\em Nature Physics}, vol.~14, pp.~303--308, Mar. 2018.

\bibitem{Zihlmann2018}
S.~Zihlmann, A.~W. Cummings, J.~H. Garcia, M.~Kedves, K.~Watanabe,
  T.~Taniguchi, C.~Schönenberger, and P.~Makk, Large spin relaxation
  anisotropy and valley-Zeeman spin-orbit coupling in
  ${\mathrm{WSe}}_{2}$/graphene/$h$-BN heterostructures {\em Phys. Rev. B},
  vol.~97, p.~075434, Feb. 2018.

\bibitem{Ghiasi2017}
T.~S. Ghiasi, J.~Ingla-Aynés, A.~A. Kaverzin, and B.~J. van Wees, Large
  Proximity-Induced Spin Lifetime Anisotropy in Transition-Metal
  Dichalcogenide/Graphene Heterostructures {\em Nano Lett.}, vol.~17,
  pp.~7528--7532, Dec. 2017.

\bibitem{Wakamura2019}
T.~Wakamura, F.~Reale, P.~Palczynski, M.~Q. Zhao, A.~T.~C. Johnson, S.~Guéron,
  C.~Mattevi, A.~Ouerghi, and H.~Bouchiat, Spin-orbit interaction induced in
  graphene by transition metal dichalcogenides {\em Phys. Rev. B}, vol.~99,
  p.~245402, June 2019.

\bibitem{Fulop2021wal}
B.~Fülöp, A.~Márffy, S.~Zihlmann, M.~Gmitra, E.~Tóvári, B.~Szentpéteri,
  M.~Kedves, K.~Watanabe, T.~Taniguchi, J.~Fabian, C.~Schönenberger, P.~Makk,
  and S.~Csonka, ``Boosting proximity spin-orbit coupling in graphene/{WSe}$_2$
  heterostructures via hydrostatic pressure.'' (in prep.).

\bibitem{Leutenantsmeyer2017}
J.~C. Leutenantsmeyer, A.~A. Kaverzin, M.~Wojtaszek, and B.~J. van Wees,
  Proximity induced room temperature ferromagnetism in graphene probed with
  spin currents {\em 2D Materials}, vol.~4, no.~1, p.~014001, 2017.

\bibitem{Karpiak2019}
B.~Karpiak, A.~W. Cummings, K.~Zollner, M.~Vila, D.~Khokhriakov, A.~M. Hoque,
  A.~Dankert, P.~Svedlindh, J.~Fabian, S.~Roche, and S.~P. Dash, Magnetic
  proximity in a van der Waals heterostructure of magnetic insulator and
  graphene {\em 2D Materials}, vol.~7, p.~015026, Dec. 2019.

\bibitem{Wang2015a}
Z.~Wang, C.~Tang, R.~Sachs, Y.~Barlas, and J.~Shi, Proximity-Induced
  Ferromagnetism in Graphene Revealed by the Anomalous Hall Effect {\em Phys.
  Rev. Lett.}, vol.~114, p.~016603, Jan 2015.

\bibitem{Ghiasi2020}
T.~S. Ghiasi, A.~A. Kaverzin, A.~H. Dismukes, D.~K. de~Wal, X.~Roy, and B.~J.
  van Wees, Electrical and Thermal Generation of Spin Currents by Magnetic
  Graphene 2020.

\bibitem{Zhong2017}
D.~Zhong, K.~L. Seyler, X.~Linpeng, R.~Cheng, N.~Sivadas, B.~Huang,
  E.~Schmidgall, T.~Taniguchi, K.~Watanabe, M.~A. McGuire, W.~Yao, D.~Xiao,
  K.-M.~C. Fu, and X.~Xu, Van der Waals engineering of ferromagnetic
  semiconductor heterostructures for spin and valleytronics {\em Sci Adv},
  vol.~3, p.~e1603113, May 2017.

\bibitem{Zhong2020}
D.~Zhong, K.~L. Seyler, X.~Linpeng, N.~P. Wilson, T.~Taniguchi, K.~Watanabe,
  M.~A. McGuire, K.-M.~C. Fu, D.~Xiao, W.~Yao, and X.~Xu, Layer-resolved
  magnetic proximity effect in van der Waals heterostructures {\em Nature
  Nanotechnology}, vol.~15, pp.~187--191, Mar. 2020.

\bibitem{Dean2013}
C.~R. Dean, L.~Wang, P.~Maher, C.~Forsythe, F.~Ghahari, Y.~Gao, J.~Katoch,
  M.~Ishigami, P.~Moon, M.~Koshino, T.~Taniguchi, K.~Watanabe, K.~L. Shepard,
  J.~Hone, and P.~Kim, Hofstadter’s butterfly and the fractal quantum Hall
  effect in moiré superlattices {\em Nature}, vol.~497, p.~598, May 2013.

\bibitem{Ponomarenko2013}
L.~A. Ponomarenko, R.~V. Gorbachev, G.~L. Yu, D.~C. Elias, R.~Jalil, A.~A.
  Patel, A.~Mishchenko, A.~S. Mayorov, C.~R. Woods, J.~R. Wallbank,
  M.~Mucha-Kruczynski, B.~A. Piot, M.~Potemski, I.~V. Grigorieva, K.~S.
  Novoselov, F.~Guinea, V.~I. Fal’ko, and A.~K. Geim, Cloning of Dirac
  fermions in graphene superlattices {\em Nature}, vol.~497, pp.~594--597, May
  2013.

\bibitem{Hunt2013}
B.~Hunt, J.~D. Sanchez-Yamagishi, A.~F. Young, M.~Yankowitz, B.~J. LeRoy,
  K.~Watanabe, T.~Taniguchi, P.~Moon, M.~Koshino, P.~Jarillo-Herrero, and R.~C.
  Ashoori, Massive Dirac Fermions and Hofstadter Butterfly in a van der Waals
  Heterostructure {\em Science}, vol.~340, p.~1427, June 2013.

\bibitem{KrishnaKumar2017}
R.~Krishna~Kumar, X.~Chen, G.~H. Auton, A.~Mishchenko, D.~A. Bandurin, S.~V.
  Morozov, Y.~Cao, E.~Khestanova, M.~Ben~Shalom, A.~V. Kretinin, K.~S.
  Novoselov, L.~Eaves, I.~V. Grigorieva, L.~A. Ponomarenko, V.~I. Fal’ko, and
  A.~K. Geim, High-temperature quantum oscillations caused by recurring Bloch
  states in graphene superlattices {\em Science}, vol.~357, p.~181, July 2017.

\bibitem{KrishnaKumar2018}
R.~Krishna~Kumar, A.~Mishchenko, X.~Chen, S.~Pezzini, G.~H. Auton, L.~A.
  Ponomarenko, U.~Zeitler, L.~Eaves, V.~I. Fal’ko, and A.~K. Geim, High-order
  fractal states in graphene superlattices {\em Proc Natl Acad Sci USA},
  vol.~115, p.~5135, May 2018.

\bibitem{Ribeiro-Palau2018}
R.~Ribeiro-Palau, C.~Zhang, K.~Watanabe, T.~Taniguchi, J.~Hone, and C.~R. Dean,
  Twistable electronics with dynamically rotatable heterostructures {\em
  Science}, vol.~361, p.~690, Aug. 2018.

\bibitem{Wang2019}
L.~Wang, S.~Zihlmann, M.-H. Liu, P.~Makk, K.~Watanabe, T.~Taniguchi,
  A.~Baumgartner, and C.~Schönenberger, New Generation of Moiré Superlattices
  in Doubly Aligned hBN/Graphene/hBN Heterostructures {\em Nano Lett.},
  vol.~19, pp.~2371--2376, Apr. 2019.

\bibitem{cao2018a}
Y.~Cao, V.~Fatemi, S.~Fang, K.~Watanabe, T.~Taniguchi, E.~Kaxiras, and
  P.~Jarillo-Herrero, Unconventional superconductivity in magic-angle graphene
  superlattices {\em Nature}, vol.~556, p.~43, Mar. 2018.

\bibitem{cao2018b}
Y.~Cao, V.~Fatemi, A.~Demir, S.~Fang, S.~L. Tomarken, J.~Y. Luo, J.~D.
  Sanchez-Yamagishi, K.~Watanabe, T.~Taniguchi, E.~Kaxiras, R.~C. Ashoori, and
  P.~Jarillo-Herrero, Correlated insulator behaviour at half-filling in
  magic-angle graphene superlattices {\em Nature}, vol.~556, p.~80, Mar. 2018.

\bibitem{Weng2014}
H.~Weng, X.~Dai, and Z.~Fang, Transition-Metal Pentatelluride
  $\mathrm{ZrTe}{}_{5}$ and $\mathrm{HfTe}{}_{5}$: A Paradigm for Large-Gap
  Quantum Spin Hall Insulators {\em Phys. Rev. X}, vol.~4, p.~011002, Jan.
  2014.

\bibitem{Munoz2016}
F.~Munoz, H.~P.~O. Collado, G.~Usaj, J.~O. Sofo, and C.~A. Balseiro, Bilayer
  graphene under pressure: Electron-hole symmetry breaking, valley Hall effect,
  and Landau levels {\em Phys. Rev. B}, vol.~93, p.~235443, June 2016.

\bibitem{Fan2017}
Z.~Fan, Q.-F. Liang, Y.~B. Chen, S.-H. Yao, and J.~Zhou, Transition between
  strong and weak topological insulator in ZrTe5 and HfTe5 {\em Scientific
  Reports}, vol.~7, p.~45667, Apr. 2017.

\bibitem{Nam2017}
N.~N.~T. Nam and M.~Koshino, Lattice relaxation and energy band modulation in
  twisted bilayer graphene {\em Phys. Rev. B}, vol.~96, p.~075311, Aug. 2017.

\bibitem{Zhang2017}
Y.~Zhang, C.~Wang, L.~Yu, G.~Liu, A.~Liang, J.~Huang, S.~Nie, X.~Sun, Y.~Zhang,
  B.~Shen, J.~Liu, H.~Weng, L.~Zhao, G.~Chen, X.~Jia, C.~Hu, Y.~Ding, W.~Zhao,
  Q.~Gao, C.~Li, S.~He, L.~Zhao, F.~Zhang, S.~Zhang, F.~Yang, Z.~Wang, Q.~Peng,
  X.~Dai, Z.~Fang, Z.~Xu, C.~Chen, and X.~J. Zhou, Electronic evidence of
  temperature-induced Lifshitz transition and topological nature in ZrTe5 {\em
  Nature Communications}, vol.~8, p.~15512, May 2017.

\bibitem{Guinea2019}
F.~Guinea and N.~R. Walet, Continuum models for twisted bilayer graphene:
  Effect of lattice deformation and hopping parameters {\em Phys. Rev. B},
  vol.~99, p.~205134, May 2019.

\bibitem{Chebrolu2019}
N.~R. Chebrolu, B.~L. Chittari, and J.~Jung, Flat bands in twisted double
  bilayer graphene {\em Phys. Rev. B}, vol.~99, p.~235417, June 2019.

\bibitem{Lin2020}
X.~Lin, H.~Zhu, and J.~Ni, Pressure-induced gap modulation and topological
  transitions in twisted bilayer and twisted double bilayer graphene {\em Phys.
  Rev. B}, vol.~101, p.~155405, Apr 2020.

\bibitem{Dave2004}
M.~Dave, R.~Vaidya, S.~G. Patel, and A.~R. Jani, High pressure effect on MoS2
  and MoSe2 single crystals grown by CVT method {\em Bulletin of Materials
  Science}, vol.~27, pp.~213--216, Apr. 2004.

\bibitem{Aksoy2006}
R.~Aksoy, Y.~Ma, E.~Selvi, M.~C. Chyu, A.~Ertas, and A.~White, X-ray
  diffraction study of molybdenum disulfide to 38.8GPa {\em Journal of Physics
  and Chemistry of Solids}, vol.~67, pp.~1914--1917, Sept. 2006.

\bibitem{Aksoy2008}
R.~Aksoy, E.~Selvi, and Y.~Ma, X-ray diffraction study of molybdenum diselenide
  to 35.9GPa {\em Journal of Physics and Chemistry of Solids}, vol.~69,
  pp.~2138--2140, Sept. 2008.

\bibitem{Hromadova2013}
L.~Hromadov\'a, R.~Marto\ifmmode~\check{n}\else \v{n}\fi{}\'ak, and E.~Tosatti,
  Structure change, layer sliding, and metallization in high-pressure
  MoS${}_{2}$ {\em Phys. Rev. B}, vol.~87, p.~144105, Apr 2013.

\bibitem{riflikova2014}
M.~Rifliková, R.~Martoňák, and E.~Tosatti, Pressure-induced gap closing and
  metallization of $\mathrm{Mo}{\mathrm{Se}}_{2}$ and
  $\mathrm{Mo}{\mathrm{Te}}_{2}$ {\em Phys. Rev. B}, vol.~90, p.~035108, July
  2014.

\bibitem{Zamborszky2001}
F.~Zamborszky, I.~Kezsmarki, L.~K. Montgomery, and G.~Mihaly, Pressure induced
  crossover in the electronic states of (TMTTF)2Br {\em null}, vol.~249,
  pp.~57--62, Jan. 2001.

\bibitem{kezsmarki2007}
I.~K\'ezsm\'arki, R.~Ga\'al, C.~C. Homes, B.~S\'{\i}pos, H.~Berger,
  S.~Bord\'acs, G.~Mih\'aly, and L.~Forr\'o, High-pressure infrared
  spectroscopy: Tuning of the low-energy excitations in correlated electron
  systems {\em Phys. Rev. B}, vol.~76, p.~205114, Nov 2007.

\bibitem{kusmartseva2009}
A.~F. Kusmartseva, B.~Sipos, H.~Berger, L.~Forró, and E.~Tutiš, Pressure
  Induced Superconductivity in Pristine
  $1T\mathrm{\text{\ensuremath{-}}}{\mathrm{TiSe}}_{2}$ {\em Phys. Rev. Lett.},
  vol.~103, p.~236401, Nov. 2009.

\bibitem{vangennep2016}
D.~VanGennep, A.~Linscheid, D.~Jackson, S.~Weir, Y.~Vohra, H.~Berger,
  G.~Stewart, R.~Hennig, P.~Hirschfeld, and J.~J. Hamlin, Pressure-induced
  superconductivity in the giant Rashba system BiTeI {\em J. Phys.: Condens.
  Matter}, 2016.

\bibitem{forro2000}
L.~Forró, R.~Gaál, H.~Berger, P.~Fazekas, K.~Penc, I.~Kézsmárki, and
  G.~Mihály, Pressure Induced Quantum Critical Point and Non-Fermi-Liquid
  Behavior in ${\mathrm{BaVS}}_{3}$ {\em Phys. Rev. Lett.}, vol.~85,
  pp.~1938--1941, Aug. 2000.

\bibitem{kezsmarki2001}
I.~Kézsmárki, S.~Csonka, H.~Berger, L.~Forró, P.~Fazekas, and G.~Mihály,
  Pressure dependence of the spin gap in ${\mathrm{BaVS}}_{3}$ {\em Phys. Rev.
  B}, vol.~63, p.~081106, Feb. 2001.

\bibitem{kezsmarki2005}
I.~Kézsmárki, G.~Mihály, R.~Gaál, N.~Barišić, H.~Berger, L.~Forró, C.~C.
  Homes, and L.~Mihály, Pressure-induced suppression of the spin-gapped
  insulator phase in $\mathrm{Ba}\mathrm{V}{\mathrm{S}}_{3}$: An infrared
  optical study {\em Phys. Rev. B}, vol.~71, p.~193103, May 2005.

\bibitem{csontos2005}
M.~Csontos, G.~Mihály, B.~Jankó, T.~Wojtowicz, X.~Liu, and J.~K. Furdyna,
  Pressure-induced ferromagnetism in (In,Mn)Sb dilute magnetic semiconductor
  {\em Nature Materials}, vol.~4, pp.~447--449, June 2005.

\bibitem{barisic2002}
N.~Barišić, R.~Gaál, I.~Kézsmárki, G.~Mihály, and L.~Forró, Pressure
  dependence of the thermoelectric power of single-walled carbon nanotubes {\em
  Phys. Rev. B}, vol.~65, p.~241403, May 2002.

\bibitem{Yankowitz2018}
M.~Yankowitz, J.~Jung, E.~Laksono, N.~Leconte, B.~L. Chittari, K.~Watanabe,
  T.~Taniguchi, S.~Adam, D.~Graf, and C.~R. Dean, Dynamic band-structure tuning
  of graphene moiré superlattices with pressure {\em Nature}, vol.~557,
  pp.~404--408, May 2018.

\bibitem{Yankowitz2019}
M.~Yankowitz, S.~Chen, H.~Polshyn, Y.~Zhang, K.~Watanabe, T.~Taniguchi,
  D.~Graf, A.~F. Young, and C.~R. Dean, Tuning superconductivity in twisted
  bilayer graphene {\em Science}, vol.~363, p.~1059, Mar. 2019.

\bibitem{Song2019}
T.~Song, Z.~Fei, M.~Yankowitz, Z.~Lin, Q.~Jiang, K.~Hwangbo, Q.~Zhang, B.~Sun,
  T.~Taniguchi, K.~Watanabe, M.~A. McGuire, D.~Graf, T.~Cao, J.-H. Chu, D.~H.
  Cobden, C.~R. Dean, D.~Xiao, and X.~Xu, Switching 2D magnetic states via
  pressure tuning of layer stacking {\em Nature Materials}, vol.~18,
  pp.~1298--1302, Dec. 2019.

\bibitem{Leutenantsmeyer2018}
J.~C. Leutenantsmeyer, J.~Ingla-Ayn\'es, J.~Fabian, and B.~J. van Wees,
  Observation of Spin-Valley-Coupling-Induced Large Spin-Lifetime Anisotropy in
  Bilayer Graphene {\em Phys. Rev. Lett.}, vol.~121, p.~127702, Sep 2018.

\bibitem{Xu2018}
J.~Xu, T.~Zhu, Y.~K. Luo, Y.-M. Lu, and R.~K. Kawakami, Strong and Tunable
  Spin-Lifetime Anisotropy in Dual-Gated Bilayer Graphene {\em Phys. Rev.
  Lett.}, vol.~121, p.~127703, Sep 2018.

\bibitem{CastroNeto2009}
A.~H. Castro~Neto, F.~Guinea, N.~M.~R. Peres, K.~S. Novoselov, and A.~K. Geim,
  The electronic properties of graphene {\em Rev. Mod. Phys.}, vol.~81,
  pp.~109--162, Jan 2009.

\bibitem{Zomer2014}
P.~J. Zomer, M.~H.~D. Guimarães, J.~C. Brant, N.~Tombros, and B.~J. van Wees,
  Fast pick up technique for high quality heterostructures of bilayer graphene
  and hexagonal boron nitride {\em Appl. Phys. Lett.}, vol.~105, p.~013101,
  July 2014.

\bibitem{Avsar2014}
A.~Avsar, J.~Y. Tan, T.~Taychatanapat, J.~Balakrishnan, G.~K.~W. Koon, Y.~Yeo,
  J.~Lahiri, A.~Carvalho, A.~S. Rodin, E.~C.~T. O’Farrell, G.~Eda, A.~H.
  Castro~Neto, and B.~Özyilmaz, Spin-orbit proximity effect in graphene {\em
  Nature Communications}, vol.~5, p.~4875, Sept. 2014.

\bibitem{Yang2016}
B.~Yang, M.-F. Tu, J.~Kim, Y.~Wu, H.~Wang, J.~Alicea, R.~Wu, M.~Bockrath, and
  J.~Shi, Tunable spin-orbit coupling and symmetry-protected edge states in
  graphene/WS 2 {\em 2D Materials}, vol.~3, p.~031012, Sept. 2016.

\bibitem{Yang2017}
B.~Yang, M.~Lohmann, D.~Barroso, I.~Liao, Z.~Lin, Y.~Liu, L.~Bartels,
  K.~Watanabe, T.~Taniguchi, and J.~Shi, Strong electron-hole symmetric Rashba
  spin-orbit coupling in graphene/monolayer transition metal dichalcogenide
  heterostructures {\em Phys. Rev. B}, vol.~96, p.~041409, Jul 2017.

\bibitem{Ringer2018}
S.~Ringer, S.~Hartl, M.~Rosenauer, T.~V\"olkl, M.~Kadur, F.~Hopperdietzel,
  D.~Weiss, and J.~Eroms, Measuring anisotropic spin relaxation in graphene
  {\em Phys. Rev. B}, vol.~97, p.~205439, May 2018.

\bibitem{Island2019}
J.~O. Island, X.~Cui, C.~Lewandowski, J.~Y. Khoo, E.~M. Spanton, H.~Zhou,
  D.~Rhodes, J.~C. Hone, T.~Taniguchi, K.~Watanabe, L.~S. Levitov, M.~P.
  Zaletel, and A.~F. Young, Spin-orbit-driven band inversion in bilayer
  graphene by the van der Waals proximity effect {\em Nature}, vol.~571,
  pp.~85--89, July 2019.

\bibitem{Amann2012.05718}
J.~Amann, T.~Völkl, D.~Kochan, K.~Watanabe, T.~Taniguchi, J.~Fabian, D.~Weiss,
  and J.~Eroms, Gate-tunable Spin-Orbit-Coupling in Bilayer
  Graphene-WSe$_2$-heterostructures 2020.

\bibitem{IhnBook2004}
T.~Ihn, {\em Electronic Quantum Transport in Mesoscopic Semiconductor
  Structures}.
\newblock Springer-Verlag New York, 2004.

\bibitem{Fulop2021wal_arxiv}
B.~Fülöp, A.~Márffy, S.~Zihlmann, M.~Gmitra, E.~Tóvári, B.~Szentpéteri,
  M.~Kedves, K.~Watanabe, T.~Taniguchi, J.~Fabian, C.~Schönenberger, P.~Makk,
  and S.~Csonka, Boosting proximity spin orbit coupling in graphene/WSe$_2$
  heterostructures via hydrostatic pressure 2021.
\newblock arXiv:2103.13325.

\bibitem{Murata1997}
K.~Murata, H.~Yoshino, H.~Yadav, Y.~Honda, and N.~Shirakawa, Pt resistor
  thermometry and pressure calibration in a clamped pressure cell with the
  medium, Daphne 7373 {\em Review of Scientific Instruments}, vol.~68,
  pp.~2490--2493, June 1997.

\bibitem{Yokogawa2007}
K.~Yokogawa, K.~Murata, H.~Yoshino, and S.~Aoyama, Solidification of
  High-Pressure Medium Daphne 7373 {\em Japanese Journal of Applied Physics},
  vol.~46, pp.~3636--3639, June 2007.

\end{thebibliography}
\bibliographystyle{ieeetr}

\end{document}